\documentclass[aps,pra,twocolumn,groupedaddress,reprint,longbibliography]{revtex4-2}

\usepackage{graphicx}
\usepackage{epstopdf}
\usepackage{comment}
\usepackage{amsmath,amssymb}
\usepackage{bm}
\usepackage{float}
\usepackage[utf8]{inputenc}
\usepackage[T1]{fontenc}
\usepackage{color}

\newcommand{\ctext}[1]{\raise0.2ex\hbox{\textcircled{\scriptsize{#1}}}}
\newcommand{\vect}[1]{{\mbox{\boldmath $#1$}}}

\begin{document}

\title{Nonreciprocal phonon propagation via spatially asymmetric magnon-phonon coupling}

\author{Hiroyuki Moriya}
\email[]{hiroyuki.moriya@ntt.com}

\affiliation{Basic Research Laboratories, NTT, Inc., Kanagawa, Japan}

\author{Motoki Asano}

\affiliation{Basic Research Laboratories, NTT, Inc., Kanagawa, Japan}

\author{Daiki Hatanaka}

\affiliation{Basic Research Laboratories, NTT, Inc., Kanagawa, Japan}

\author{Yoshitaka Taniyasu}

\affiliation{Basic Research Laboratories, NTT, Inc., Kanagawa, Japan}

\author{Hajime Okamoto}

\affiliation{Basic Research Laboratories, NTT, Inc., Kanagawa, Japan}

\author{Hiroshi Yamaguchi}

\affiliation{Basic Research Laboratories, NTT, Inc., Kanagawa, Japan}

\date{\today}

\begin{abstract}
Nonreciprocal propagation of surface acoustic waves~(SAWs) based on the spatial asymmetry of magnon–phonon coupling is demonstrated. This nonreciprocity is enabled by an acoustic wavelength scale thick magnetic layer formed under a thin piezoelectric film. In this configuration, magnon modes activated by dipole–dipole interactions are localized near either the top or bottom interface depending on the propagation direction of the SAW. As a result, the spatial overlap between interfacial magnon and surface phonon modes is expected to become direction dependent, in a manner that leads to distinct unidirectional propagation of SAWs. Notably, the resulting nonreciprocity reaches the highest level among those reported for SAW devices based on a single magnetic layer. This finding will establish a new strategy of nonreciprocal acoustic propagation in a structurally simple magnomechanical device.
\end{abstract}

\maketitle

\section{Introduction}

In recent years, magnetoelastic devices combining ferromagnetic films with surface acoustic wave~(SAW) structures have attracted considerable interest as platforms for unconventional SAW control. The magnetoelastic coupling achieved by these devices enables nonreciprocal SAW propagation, which offers functionalities such as microwave signal routing and isolation~\cite{weiler2011elastically,delsing20192019,li2021advances,will2022tutorial,puebla2022perspectives,luo2024magnetoelectric}. Although SAWs themselves are time-reversal symmetric, coupling to magnons under a magnetic field breaks this symmetry and results in direction-dependent SAW transmission~\cite{shah2020giant,kuss2022chiral}.
So far, two main mechanisms have been used to achieve nonreciprocal phonon propagation in SAW-based magnetoelastic systems. The first exploits the chirality of Rayleigh-type SAWs and magnon spin precession~\cite{dreher2012surface, sasaki2017nonreciprocal,xu2020nonreciprocal}.
Since the lattice rotation of the SAW reverses with reversed propagation direction, the magnetoelastic coupling becomes direction dependent, and that dependency leads to nonreciprocal phonon transmission~\cite{weiler2011elastically,dreher2012surface,sasaki2017nonreciprocal,tateno2020highly,hernandez2020large}. This mechanism can be realized in simple single-layer structures, but the resulting nonreciprocity is typically small, in the order of a few dB/mm~\cite{sasaki2017nonreciprocal}.
The second mechanism arises from asymmetric magnon dispersion generated by dipole-dipole interaction in magnetic heterostructures \cite{shah2020giant,kuss2021nonreciprocal,matsumoto2022large,kuss2023nonreciprocal,kuss2023giant}.
In this case, the magnon-phonon resonance frequency depends on the propagation direction, and that dependency produces different resonance dips for forward and backward propagation and results in strong nonreciprocal transmission at a given magnetic field~\cite{verba2018nonreciprocal,kuss2023nonreciprocal}. Although large nonreciprocity exceeding 250~dB/mm has been reported~\cite{kuss2023giant}, this mechanism generally requires complex, precisely fabricated multilayer structures.

Beyond such interactions, magnons themselves can exhibit nonreciprocal propagation. When the in-plane equilibrium magnetization of the ferromagnetic film is perpendicular to the magnon wavevector, surface magnons known as Damon-Eshbach~(DE) modes emerge~\cite{damon1961magnetostatic}. In films with thickness comparable to the magnon wavelength, DE modes become localized near either the top or bottom surface of the magnetic layer depending on the propagation direction, and that localization results in intrinsically nonreciprocal magnon transport. Although such surface asymmetry is expected to enhance nonreciprocal propagation of SAWs, its potential has remained unexplored.

\begin{figure}[htbp]
    \centering
    \includegraphics[scale=1]{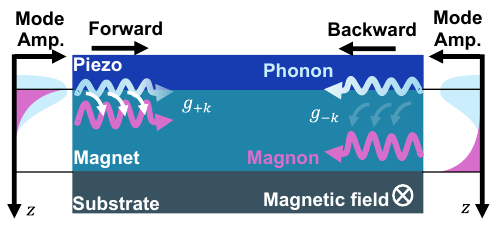}
    \caption{Schematic of asymmetric magnomechanical coupling between phonon and magnon propagation with a mode amplitude (amp.) in the thickness ($z$) direction.}
    \label{fig0}
\end{figure}

In this study,  SAW propagation with a large degree of nonreciprocity in a structurally simple piezo-on-magnet (POM) system consisting of a single 5.7-$\mathrm{\mu m}$-thick Y$_3$Fe$_5$O$_{12}$~(YIG) layer is demonstrated. In this configuration, the shear strain associated with the SAW is strongly confined near the piezoelectric/magnetic interface~(see Fig.~\ref{fig0}), and that confinement makes the magnon-phonon coupling sensitive to the spatial profile of the magnon mode across the film thickness. As a result, surface-localized magnon modes propagating in opposite directions can exhibit different spatial overlaps with the phonons, and this status leads to asymmetric coupling strengths. Combined with the strong magnon-phonon coupling previously reported in POM structures~\cite{hatanaka2025chip}, this mechanism enables pronounced nonreciprocal phonon absorption.

\section{Experiment}

\begin{figure}[htbp]
    \centering
    \includegraphics[scale=1]{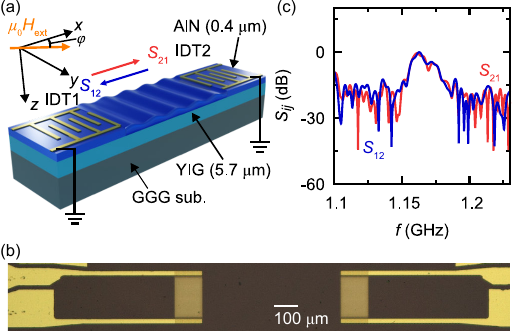}
    \caption{(a) Schematic illustration of a surface acoustic wave (SAW) device containing Y$_3$Fe$_5$O$_{12}$~(YIG) on a Gd$_3$Ga$_5$O$_{12}$~(GGG) substrate, covered by piezoelectric AlN film. (b) Photograph of the SAW device. The two interdigitated transducers (IDTs) are connected to the electrode pads, and the SAW propagates between the IDTs.(c) Representative SAW transmission spectra for $S_{12}$ (blue) and $S_{21}$ (red) in the absence of a magnetic field.}
    \label{fig1}
\end{figure}

The SAW device and its coordinate axes are shown schematically in Fig.~\ref{fig1}(a).
The device was fabricated on an AlN(0.4~$\mathrm{\mu m}$)/YIG(5.7~$\mathrm{\mu m}$)/Gd$_3$Ga$_5$O$_{12}$~(GGG) substrate and consists of a pair of Ti(5~nm)/Au(50~nm) inter-digit transducers~(IDTs) located 640~$\mathrm{\mu m}$ apart as shown in Fig.~\ref{fig1}(b). The YIG film is made much thicker than in previous studies~\cite{ba2025nonreciprocal, kunz2025efficient} to enhance the separation of localized magnon modes. The width and intervals of the IDT fingers are set to 0.8~$\mathrm{\mu m}$. The piezoelectric AlN layer allows SAWs to be electrically generated and detected at the IDTs, whereas the magnetic YIG layer lies below the AlN film and interacts with the SAWs as they propagate. The SAW transmission was characterized by measuring the forward and backward scattering parameters with a vector network analyzer~(VNA).
 $S_{ij}$ is defined as the amplitude of the scattering parameters for waves propagating from port $j$ to port $i$ normalized by the value at $\mu_0 H_{\rm ext}=10$~mT. All measurements were carried out at room temperature with the input radio-frequency (RF) power fixed at 10~dBm. Typical SAW forward ($S_{21}$) and backward ($S_{12}$) transmission spectra in the absence of a magnetic field are shown in Fig.~\ref{fig1}(c). The SAW itself is reciprocal with respect to the propagation direction. From these spectra, the center frequency was determined to be 1.16~GHz and bandwidth was determined to be 17.7~MHz. 

\begin{figure}[tbp]
    \centering
    \includegraphics[scale=1]{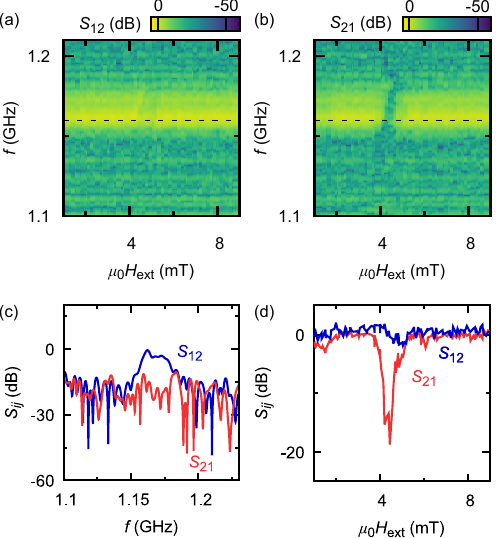}
    \caption{(a)-(b) Measured SAW spectra as a function of external magnetic field $\mu_0H_{\rm ext}$ and frequency $f$ at angle $\phi=90^\circ$ for (a) $S_{12}$ and (b) $S_{21}$. (c) Transmission SAW spectrum for $S_{12}$ (blue) and $S_{21}$ (red) of the SAW device around the first harmonic frequency measured at $\mu_0H_{\rm ext}=4.2$~mT. (d) Magnetic-field dependence of SAW absorption signal $S_{ij}$ measured at $f=1.16$~GHz for $S_{12}$ (blue) and $S_{21}$ (red) at $\phi=90^\circ$.}
    \label{fig2}
\end{figure}

To probe the magnon--phonon--induced nonreciprocity, an in-plane external magnetic field $\mu_0H_{\rm ext}$ was swept from $10$ to $0$~mT with the magnetic field oriented at $\phi=90^\circ$ relative to the propagation direction (see Fig.~\ref{fig1}(a)). Color maps of the backward and forward SAW transmission, $S_{12}$ and $S_{21}$, plotted as functions of external magnetic field $\mu_0H_{\rm ext}$ and excitation frequency are respectively shown in Figs.~\ref{fig2}(a) and \ref{fig2}(b). A pronounced absorption dip in $S_{21}$ is clearly visible, whereas such a dip in $S_{12}$ is obscured. This dip reflects the resonant conversion of acoustic energy into magnetic excitations via magnon-phonon coupling. When the SAW frequency matches the resonance frequency of the field-tuned magnon mode, the SAW acoustic power is transferred to the magnon and dissipated through magnetic damping. As a result, transmission amplitude is reduced. The asymmetry between $S_{21}$ and $S_{12}$ indicates that the magnetoelastic interaction depends on the direction of propagation of the SAW.

 Frequency spectra of $S_{12}$ and $S_{21}$ taken at $\mu_0H_{\rm ext}=4.2$~mT are shown in Fig.~\ref{fig2}(c).  The SAW transmission spectrum for $S_{12}$ identified in Fig.~\ref{fig1}(c) remains clearly visible, whereas that for $S_{21}$ disappears under the same magnetic field configuration. This finding indicates that the SAW transmission differs markedly between opposite propagation directions.
 The magnetic field dependence of $S_{12}$ and $S_{21}$, extracted at the center frequency of the SAW mode indicated by the dashed lines in Figs~\ref{fig2}(a) and (b), is shown in Fig~\ref{fig2}(d). While $S_{12}$ is mostly insensitive to variation in $\mu_0 H_{\rm ext}$, $S_{21}$ pronouncedly dips around $\mu_0 H_{\rm ext}=4.2$~mT. This clear contrast in the behavior of $S_{12}$ and $S_{21}$ demonstrates that the disappearance of the SAW transmission spectrum in Fig.~\ref{fig2}(c) originates from strong resonant absorption occurring in the forward propagation direction of the SAW.
The absorption contrast $\Delta S = |S_{12} - S_{21}|$ (logarithmic scale) clearly demonstrates a large nonreciprocal response exceeding $\Delta S=16.1$~dB at 4.2~mT.
After normalization by the IDT separation $L=0.64$~mm, this value corresponds to a nonreciprocal attenuation of $\Delta S/L=25.2$~dB/mm.
The nonreciprocity is among the largest reported for SAW devices based on a single magnetic layer and cannot be explained primarily by the two conventional mechanisms discussed in the Introduction: chirality mismatch and asymmetric magnon dispersion. First, we consider the chirality mismatch mechanism between the SAW lattice rotation and magnon precession. For the magnetic field orientations close to $\phi=90^\circ$, where the SAW wavevector is perpendicular to the equilibrium magnetization, this mechanism is known to produce only weak nonreciprocity~\cite{sasaki2017nonreciprocal}. Second, the contribution of asymmetric magnon dispersion to the nonreciprocity is also limited. If the nonreciprocity originated from the asymmetric magnon dispersion, $S_{12}$ would also be expected to dip near the magnetic field at which $S_{21}$ dips. However, as shown in Fig.~\ref{fig2}(d), $S_{12}$ does not dip in this manner. The absence of a corresponding $S_{12}$ dip and the pronounced attenuation observed only in $S_{21}$ at $\phi=90^\circ$ indicate that neither chirality mismatch nor asymmetric magnon dispersion can account for the observed nonreciprocity.
Instead, the absorption contrast, $\Delta S=16.1$~dB, can be theoretically reproduced by taking into account the magnon–phonon coupling with a surface-localized magnon mode (see Appendices B and C). In the model, the surface-localized magnon mode is assumed to have an exponential profile along the thickness ($z$) direction with a penetration depth $\delta$, which is approximately given by $\delta\sim\lambda_a$, where $\lambda_a$ is the phonon wavelength~\cite{damon1961magnetostatic}. With the device geometry and loss factors referenced from previous work~\cite{hatanaka2025chip}, the model yields an absorption contrast of $\Delta S=17.1$~dB, which agrees well with the experimental value. This agreement suggests that the large nonreciprocity is mainly caused by spatially asymmetric magnon-phonon coupling involving a surface-localized magnon mode.

\begin{figure}[htbp]
    \centering
    \includegraphics[scale=1]{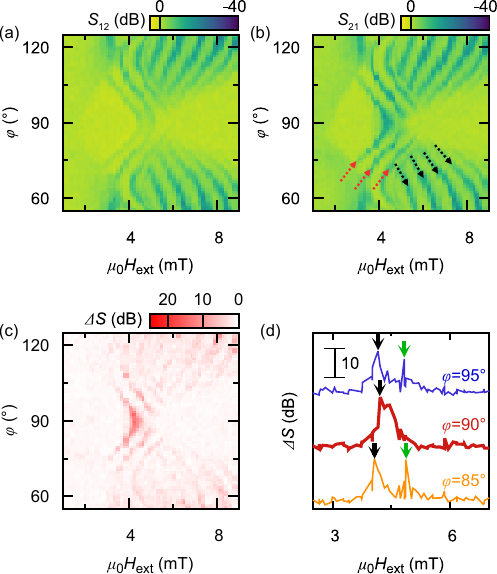}
    \caption{(a)-(b) Color maps of SAW absorption signals as functions of external magnetic field $\mu_0H_{\rm ext}$ and its in-plane angle $\phi$ for (a) $S_{12}$ and (b) $S_{21}$. (c) Color map of the nonreciprocity defined as $\Delta S = |S_{12}-S_{21}|$. (d) Line-cut profiles of $\Delta S$ extracted from panel (c) at $\phi=85^\circ$ (orange), $90^\circ$ (red) and $95^\circ$ (blue).}
    \label{fig3}
\end{figure}

In addition to the surface-localized magnon mode discussed above, other magnon modes may also contribute to the nonreciprocal phonon absorption because the YIG layer is thicker than the magnon wavelength~\cite{hatanaka2025chip}.
To investigate multimode magnon contributions to nonreciprocal phonon absorption, we examined the magnetic-field-angle dependence of the transmission signals. Color maps of $S_{12}$ and $S_{21}$ as functions of external magnetic field $\mu_0H_{\rm ext}$ and in-plane field angle $\phi$ are shown in Figs.~\ref{fig3}(a) and \ref{fig3}(b). Multiple absorption branches emerge around $\phi = 90^\circ$ and indicate magnetoelastic coupling between the propagating SAWs and several spin-wave modes supported by the thick YIG film. Absorption contrast $\Delta S$, evaluated from Figs.~\ref{fig3}(a) and \ref{fig3}(b) and shown in Fig.~\ref{fig3}(c), allows us to classify the resonant branches into reciprocal (black arrows in Fig.~\ref{fig3}(b)) and nonreciprocal (red arrows in Fig.~\ref{fig3}(b)) modes. The reciprocal branches shift toward higher resonance fields as $\phi$ deviates from $90^\circ$ in a manner that is consistent with the angular evolution expected for conventional bulk spin-wave modes governed by dipolar interactions~\cite{kalinikos1986theory,serga2010yig}.
In contrast, the nonreciprocal branches exhibit a systematic shift toward lower resonance fields as the field angle deviates from $90^\circ$. Such a negative field shift suggests the presence of an additional effective anisotropy field that reduces the internal field acting on the magnetization of the YIG layer. Notably, this shift appears exclusively in the nonreciprocal modes, implying that it is intrinsically linked to the surface-localized character of the modes. The angular evolution of the nonreciprocal response is further clarified in Fig.~\ref{fig3}(d), which shows line-cut profiles of $\Delta S$. The nonreciprocity reaches its maximum at $\phi = 90^\circ$.
As $\phi$ deviates from $90^\circ$, the dominant nonreciprocal peak is suppressed (see black arrows in Fig.~\ref{fig3}(d)), while another peaks appear at a different resonance field (see green arrows in Fig.~\ref{fig3}(d)). The emergence of this secondary peaks indicate the participation of multiple surface-localized spin-wave modes in the observed nonreciprocal absorption. This finding implies that the magnitude and spectral profile of the nonreciprocity can be tuned by controlling the contribution of different magnon modes.

The magnon mode is localized near the top or bottom surface of the YIG layer according to the sign of the wave vector. As a result, magnon modes are highly sensitive to surface anisotropy. Interface-induced anisotropy can arise from strain and symmetry breaking at the magnetic–piezoelectric boundary. In the present AlN/YIG heterostructure, the lattice constant of YIG (12.4~\AA~\cite{wang2017comparative}) differs substantially from that of AlN (wurtzite lattice constants $a = 3.11$~\AA\ and $c = 4.98$~\AA~\cite{ruiz1994electronic}), and that difference gives rise to interfacial strain and magnetoelastic contributions to the effective surface anisotropy. This strain-induced anisotropy locally modifies the internal field at the top AlN/YIG interface and influences the magnon modes, and this modification explains the observed negative resonance-field shift.
Although the directional difference between $S_{12}$ and $S_{21}$ indicates that nonreciprocal phonon absorption mainly originates from the asymmetrically localized magnon modes, the field-angle dependence suggests that the properties of the interfaces also play an additional role in inducing the nonreciprocity. To clarify the microscopic origin of the observed pronounced nonreciprocity, detailed studies, including structural characterization of the interfaces, are required.

\section{Conclusion}
Pronounced nonreciprocal propagation of SAWs in a POM AlN/YIG/GGG structure, where a thick magnetic layer can be integrated without requiring complex multilayer engineering, was demonstrated. A structurally simple SAW device consisting of a single magnetic layer achieved nonreciprocity of 25.2~dB/mm, demonstrating the potential for realizing large nonreciprocal responses without using magnetic multilayers.
The nonreciprocity clearly depends on both the magnitude and sign of the external magnetic field and on in-plane field angle $\phi$, and it reaches its maximum at $\phi=90^\circ$. This study aimed to utilize the asymmetry associated with DE modes to observe propagation-direction-dependent nonreciprocal behavior. While the detailed nature of the surface modes could not be fully identified, the nonreciprocity supports the interpretation that it arises from asymmetric spatial overlap between the SAW modes and surface-localized magnon modes.
Furthermore, the observed nonreciprocity is not limited to a single magnon branch. Instead, multiple magnon modes participating in the magnon-phonon hybridization display direction-dependent absorption, suggesting that the nonreciprocity is rooted in the intrinsic nature of the magnetic-excitation spectrum. 
The use of a thick single magnetic layer to create direction-dependent magnon-phonon mode overlap is distinct from previously reported strategies that employ synthetic antiferromagnets, interfacial Dzyaloshinskii–Moriya interactions, or engineered multilayer asymmetry to achieve nonreciprocal transport~\cite{shah2020giant,kuss2021nonreciprocal,matsumoto2022large,kuss2023nonreciprocal,kuss2023giant}. It thus enables the use of high-quality, low-damping, thick magnetic films while keeping the SAW response tunable by magnetic field and the device structure simple.
This strategy of using direction-dependent magnon-phonon mode overlap can be applied to a wide range of platforms supporting guided acoustic modes, piezoelectric actuation and detection, and interfacial magnetoelastic interactions. It will therefore create opportunities for field-tunable nonreciprocal magnomechanical devices such as isolators and reconfigurable on-chip signal-routing elements.

\section*{ACKNOWLEDGMENTS}
This work was supported by JSPS KAKENHI Grants, Numbers JP21H05020, JP23H05463, JP24H02235 and JP26K01424. 

\section*{Appendix}
\appendix

\section{Model}
Nonreciprocal phonon propagation can be modeled by the equations of motion derived from the following interaction Hamiltonian, which is obtained from the linearized Landau--Lifshitz--Gilbert equation and the acoustic wave equation (see Appendix~\ref{appendix:motion}):
\begin{align}
    \mathcal{H}_\mathrm{int}
    =\int \mathrm{d}k \sum_{n} g_{k,n}\left(a_k m^{\ast}_{k,n}+\mathrm{c.c.}\right),
    \label{eq:Hint}
\end{align}
where $a_k$ and $m_{k,n}$ denote the complex amplitudes of the guided phonon and magnon modes, respectively, and $g_{k,n}$ is the coupling constant between the modes.
It is assumed that both the magnon and phonon modes can be approximated as plane waves in the $x$--$y$ plane, while they possess spatial profiles in the thickness direction $z$, specified by mode number $n$.
Under these assumptions, the equations of motion are given as
\begin{align}
    \dot{a}_k=&\left(i\omega_k-\frac{\kappa_a}{2}\right)a_k
+i\sum_n g_{k,n}m_{k,n}+a_\mathrm{in}.,\label{eq:dota}\\
    \dot{m}_{k,n}=&\left(i\Omega_{k,n}-\frac{\kappa_\mathrm{m}}{2}\right)m_{k,n}
+ig_{k,n}a_k ,.\label{eq:dotm}
\end{align}
If the on-resonance condition $\omega_a=\Omega_{k,n}$ and continuous-wave excitation are assumed, mode-dependent phonon transmittance is given as
\begin{align}
    T_{k,n}=\left(\frac{1}{1+C_{k,n}}\right)^2,
\end{align}
where $C_{k,n}\equiv 4g^2_{k,n}/(\kappa_a\kappa_m)$ defines the cooperativity (see Appendix B).

The relation $g_{k,n}\neq g_{-k,n}$ holds when the magnon modes exhibit surface localization. This trend is because dipole--dipole interactions cause several magnon modes to localize at either the top or bottom surface depending on the propagation direction (i.e., the sign of $k$), whereas the phonon mode is always localized at the top surface. This asymmetric coupling directly implies $\alpha_{+}\neq\alpha_{-}$, where $\alpha_\pm\equiv \alpha_{\pm k_0,n_0}$ for arbitrary mode index $n_0$. The asymmetric magnon--phonon coupling in our system is shown schematically in Fig.~\ref{fig0}. To enhance this asymmetry, the magnetic field is oriented perpendicular to the phonon propagation direction, which is favorable for activating the dipole--dipole interaction. For this in-plane perpendicular magnetic field configuration, the magnon--phonon coupling arises from magnetostriction due to the shear strain component $\epsilon_{yz}$, which provides a non-dominant but finite contribution~\cite{weiler2011elastically,dreher2012surface}.

\section{Derivation of equation of motion}
\label{appendix:motion}
The interaction Hamiltonian for guided phonon and magnon modes is derived as follows. The derivation starts from the standard Landau--Lifshitz--Gilbert~(LLG) equation~\cite{landau1935theory,gilbert2004phenomenological} given as
\begin{align}
\frac{d\vect{M}}{dt}
=
-\gamma \vect{M}\times\mu_0\vect{H}_\mathrm{eff}
-\alpha \vect{M}\times(\vect{M}\times \mu_0\vect{H}_\mathrm{eff}),
\end{align}
where $\gamma$ is the gyromagnetic ratio and $\vect{M}$, $\vect{H}_\mathrm{eff}$, and $\mu_0$ denote magnetization vector, effective magnetic field, and vacuum magnetic permeability, respectively.

The effective magnetic field can be decomposed as
\begin{align}
\mu_0 \vect{H}_\mathrm{eff}
&=
\mu_0 \vect{H}_\mathrm{magnon}
+
\mu_0 \vect{H}_\mathrm{mm}, \\
\mu_0 \vect{H}_{\mathrm{mm},i}
&=
-\frac{1}{M_s}\frac{\partial \mathcal{E}_\mathrm{mm}}{\partial M_i}, \\
\mathcal{E}_\mathrm{mm}
&=
\sum_{i,j=\{x,y,z\}}
(b_1\delta_{ij}+b_2(1-\delta_{ij}))
\epsilon_{ij}M_iM_j,
\end{align}
where $M_s$ is the saturation magnetization, $\epsilon_{ij}$ denotes the strain tensor components, and $b_1$ and $b_2$ are the magnetoelastic coefficients corresponding to the longitudinal and shear strain components, respectively. Note that $\vect{H}_\mathrm{magnon}$ includes contributions from the external magnetic field, dipole--dipole interactions, exchange interactions, and magnetic anisotropy~\cite{dreher2012surface}.

The magnetization vector is considered to deviate slightly around the equilibrium direction, $\vect{M}=M_s\vect{e}_\mathrm{eq}+\tilde{\vect{m}}$ according to $|\tilde{\vect{m}}|\ll M_s$. A plane-wave solution for magnon modes is assumed as
\begin{align}
\tilde{\vect{m}}=&\tilde{\vect{m}}_{k,l}\Phi_{k,l}(z)e^{-i\tilde{\Omega}_{k,l}t+ikx},
\label{eq:tildevecm}
\end{align}
where $\Phi(z)$ is the magnon-mode profile in the thickness ($z$) direction. 

To discuss the magnon--phonon coupling to magnon modes that exhibit asymmetric spatial profiles due to dipole--dipole interactions, an in-plane magnetic field applied at $90^\circ$ is considered as follows. Under this field configuration, the magnetoelastic coupling originates from shear-strain component $\epsilon_{yz}$. Recently, the contribution of $\epsilon_{yz}$ to magnon-phonon coupling has been investigated both experimentally~\cite{chen2025twofold} and theoretically~\cite{kawada2025model}; therefore, it is reasonable to focus on this component, which is given as
\begin{align}
\epsilon_{yz}
=
ik u_k(t)\,\Psi_k(z)e^{-i\Omega_k t+ikx}
+\mathrm{c.c.},
\label{eq:epsilonyz}
\end{align}
where $u_k(t)$ is the acoustic displacement amplitude and $\Psi(z)$ is the spatial profile of the phonon mode. The magnomechanical coupling constant is proportional to $\sin\phi$, where $\phi$ is the in-plane magnetic-field angle.

The linearized LLG equation can then be rewritten as a set of linear coupled-mode equations for the magnon modes characterized by propagation wavenumber $k$ and discrete thickness-mode index $l$ \cite{kalinikos1986theory},
\begin{align}
\dot{\tilde{\vect{m}}}_{k,l}=i\tilde{\Omega}_{k,l}\tilde{\vect{m}}_{k,l}+\sum_{j\neq l} G_{k,lj}\tilde{\vect{m}}_{k,j}+\tilde{\vect{g}}_{k,l} u_{k,l},
\label{eq:dotmkl}
\end{align}
where $\tilde{\vect{m}}_{k,l}$ is a two-dimensional vector describing the magnon mode, and $u_l$ represents the displacement amplitude of the guided phonon mode. In Eq.~(\ref{eq:dotmkl}), $\tilde{\Omega}_{k,l}$, $G_{k,lj}$, and $\tilde{\vect{g}}_l$ represent the magnon frequency, the magnon--magnon coupling constant between different thickness modes, and the magnon--phonon coupling vector, respectively.
The components of $\tilde{\vect{g}}_l$ depend on the spatial mode overlap between the magnon and phonon profiles in the thickness direction and on the dominant strain components involved in the magnetoelastic interaction. While $\omega_{k,l}$ is mainly determined by instantaneous contributions to spin precession, such as the external magnetic field and exchange interaction, $G_{k,lj}$ is primarily governed by dipole--dipole interactions that generate intermode coupling between different thickness modes.

To make Eqs.~(\ref{eq:tildevecm})--(\ref{eq:dotmkl}) more transparent, they are subjected to the following two transformations. First, the magnon sector is partially diagonalized as
\begin{align}
\tilde{\vect{m}}_{k,l}=\sum_n m_{k,n}\vect{e}_{k,n},
\end{align}
where $\vect{e}_{k,n}$ denotes the eigenvector of the dipole-exchange magnon modes. This transformation introduces a new magnon basis $m_{k,n}$, which can exhibit surface-localized mode profiles in the thickness direction due to dipole--dipole interactions.

Second, the phonon displacement amplitude is written as
\begin{align}
u_k=\frac{1}{\sqrt{2}}\left(a_k+a_k^\ast\right),
\end{align}
where $a_k$ is the complex phonon amplitude.

After these transformations, the equation of motion for the magnon eigenmodes becomes
\begin{align}
\dot{m}_{k,n}
=
\left(i\Omega_{k,n}-\frac{\kappa_\mathrm{m}}{2}\right)m_{k,n}
+
ig_{k,n}a_k,
\end{align}
where $\Omega_{k,n}$ and $\kappa_\mathrm{m}$ are the eigenfrequency and damping rate of the magnon eigenmode, respectively. Magnon--phonon coupling constant $g_{k,n}$ is defined such that it appears in a symmetric form in the acoustic equation of motion~\cite{hatanaka2022chip}
\begin{align}
\dot{a}_k
=
\left(i\omega_k-\frac{\kappa_a}{2}\right)a_k
+
i\sum_n g_{k,n}m_{k,n}
+
a_\mathrm{in}.
\end{align}
It is explicitly given as
\begin{align}
g_{k,n}
=b_2 k \sin\phi
\sqrt{\frac{\gamma M_s}{2\omega_k \rho_0}}
\,
\frac{\int dz\, \Phi_{k,n}\Psi_k}
{\sqrt{\left(\int dz\, |\Phi_{k,n}|^2\right)
\left(\int dz\, |\Psi_k|^2\right)}},
\end{align}
where $\rho_0$ is the mass density of YIG.
The equations of motion can thus be derived from the interaction Hamiltonian given in Eq.~(\ref{eq:Hint}), which takes the similar form commonly used in waveguide optomechanics models~\cite{zoubi2016optomechanical,zhang2023quantum}.

\begin{figure}[htbp]
    \centering
    \includegraphics[scale=1]{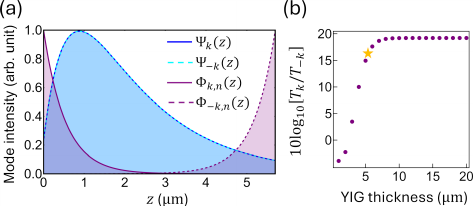}
    \caption{(a) Magnon and phonon spatial profiles in the thickness, $z$, directions. (b) Transmission ratio between the forward and backward proapagations with respect to YIG thickness. The yellow star corresponds to the experimental condition.}
    \label{figS1}
\end{figure}

The main difference between the present calculation and the calculation used in previous work~\cite{hatanaka2025chip} is that the present calculation explicitly includes the dipole--dipole interaction. Importantly, this inclusion results in $\Phi_{k,n}(z)\neq \Phi_{-k,n}(z)$ because the magnon eigenmodes exhibit asymmetric thickness profiles similar to DE surface modes \cite{damon1961magnetostatic}, whereas $\Psi_k(z)=\Psi_{-k}(z)$ holds for the phonon modes. The spatial mode profiles of the magnon modes, $\Phi_{k,n}(z)$ (purple solid) and $\Phi_{-k,n}(z)$ (purple dashed), and the phonon modes, $\Psi_k(z)$ (blue solid) and $\Psi_{-k}(z)$ (cyan dashed), calculated for our YIG structure (thickness $5.7~\mathrm{\mu}\mathrm{m}$) are shown in Fig.~\ref{figS1}(a). It is assumed that the magnon surface modes exhibit an exponential penetration of the form $\exp(-|k|z)$. As a result, the magnon--phonon mode overlap becomes strongly asymmetric between forward ($k$) and backward ($-k$) propagation.

\section{Derivation of phonon transmittance}
\label{appendix:transmittance}
Solving Eqs.~(\ref{eq:dota}) and (\ref{eq:dotm}) in the Fourier domain gives
\begin{align}
m_{k,n}
&=
\frac{i g_{k,n}}
{-i(\omega-\Omega_{k,n})+\frac{\kappa_m}{2}}
a_k ,
\\
a_k
&=
\left[
-i(\omega-\omega_k)
+\frac{\kappa_a}{2}
+
\sum_{n}
\frac{g^2_{k,n}}
{-i(\omega-\Omega_{k,n})+\frac{\kappa_m}{2}}
\right]^{-1}
a_\mathrm{in}.
\end{align}

The phonon transmittance for a given magnon mode $n$, i.e., $\omega_k=\Omega_{k,n}$, is defined as
\begin{align}
T_{k,n}\equiv\frac{4|a_k(\omega=\omega_k)|^2}{\kappa_a^2|a_\mathrm{in}|^2}.
\end{align}
Substituting the resonant condition $\omega=\omega_k=\Omega_{k,n}$ gives
\begin{align}
T_{k,n}=\left(
\frac{1}{1+\frac{4g_{k,n}^2}{\kappa_m\kappa_a}}
\right)^2=\left(\frac{1}{1+C_{k,n}}\right)^2,
\end{align}
where $C_{k,n}=4g_{k,n}^2/(\kappa_m\kappa_a)$ is the cooperativity of magnon-phonon coupling.

YIG thickness is varied under the assumption of exponentially localized surface magnon modes, and the ratio $T_k/T_{-k}$ is calculated as a measure of nonreciprocity (see Fig.~\ref{figS1}(b)). In the calculation, the loss rates $\kappa_a/2\pi=0.01~\mathrm{GHz}$, estimated from the IDT bandwidth, and $\kappa_m/2\pi=0.006~\mathrm{GHz}$ referenced from previous work using the same YIG sample~\cite{hatanaka2025chip} are used. The ratio saturates at a thickness of approximately $6$--$7~\mu\mathrm{m}$, indicating that the magnon--phonon mode overlap becomes effectively separated between the top and bottom surfaces of the YIG film. The experimental thickness of the YIG ($5.7~\mu\mathrm{m}$) is therefore close to optimal for achieving a high transmission ratio and strong nonreciprocity. Moreover, the experimentally achieved ratio ($\Delta S$) of 16.1~dB agrees well with the theoretical estimate of 17~dB.

\bibliography{00_reference}

\end{document}